**Distributed Fiber-Optic Sensing based Single-Lane Abnormal Event Detection in Low-Density Traffic Flow**

**Hemant Prasad**
Researcher
Email: h-prasad27@nec.com

**Yoshiyuki Yajima**
Researcher

**Daisuke Ikefuji**
Senior Researcher

**Takemasa Suzuki**
Manager

**Shin Tominaga**
Manager

**Hitoshi Sakurai**
Senior Manager

**Manabu Otani**
Manager

Word Count: 6588 words + 3 tables (750 words per table) = 7338 words

*H. Prasad, Y. Yajima, D. Ikefuji, T. Suzuki, S. Tominaga, H. Sakurai, and M. Otani*

**ABSTRACT**
Distributed fiber-optic sensing (DFOS) based traffic flow monitoring systems are a cost-effective wide-area traffic monitoring solution that utilize existing fiber infrastructure along roads. These systems analyse vehicle vibrations and measure average traffic speeds to detect traffic events. However, these systems face difficulties in detecting early signs of non-recurring traffic congestions in low-density traffic flow caused by presence of single-lane abnormal events. This is because average traffic speeds do not decrease quickly in such events. During abnormal events, multiple vehicles perform spontaneous braking and abrupt lane-changes to avoid obstacles on travel lanes. These vehicle behaviours gradually lead to traffic congestion. Thus, frequent lane-change maneuver, performed by multiple vehicles at similar location, may suggest occurrence of congestion-inducing abnormal events. This paper discusses methods to identify and locate occurrence of these single-lane abnormal events by detecting frequent lane-change maneuver along road sections. We first propose a method to locate vehicle positions along a road section and estimate the vehicle path. We then propose a method to detect vehicle lane-change maneuver by monitoring variations in spectral centroid of vehicle vibrations. The evaluation of our proposed methods with real traffic data for two different expressways showed 81.5% accuracy for individual vehicle path tracking and 83.5% accuracy in lane-change event detection. These results suggest that the proposed method has potential for detecting occurrence of single-lane abnormal events in low-density traffic flow so that necessary mitigation measures can be initiated before onset of traffic congestions.

H. Prasad, Y. Yajima, D. Ikefuji, T. Suzuki, S. Tominaga, H. Sakurai, and M. Otani

**INTRODUCTION**

In urban transportation, it is necessary to integrate transportation infrastructure with suitable ICT solutions to develop intelligent transportation systems (ITS) that focus on sustainability, integration, safety, and responsiveness *(1)*. Traffic congestion is one of the problems faced by ITS because it leads to increased travel time, reduced fuel efficiency and increased emissions. Thus, it is crucial for road operators and decision makers to quickly identify and quantify traffic congestions to initiate prompt mitigation strategies. Traffic congestions can be broadly classified as recurring and non-recurring types *(2)*. The recurring traffic congestion can be promptly detected by analyzing historical data and identifying congestion trends that occur in a cyclical manner. The non-recurring traffic congestion, caused either by traffic accidents, vehicle breakdowns, blocking of travel lanes, potholes, or fallen debris in travel lanes, i.e. obstruction on roads, are often uncertain and therefore, relatively difficult to identify and locate. Still, prompt detection of uncertain events where it causes traffic to move slowly in all lanes is possible. It is crucial to employ suitable traffic monitoring systems and sensor technologies to detect uncertain events. The widely adopted conventional traffic monitoring systems, which employ traffic cameras and inductive loop detectors, have issues due to shorter monitoring range and sparsity in traffic data for detecting traffic anomalies. Therefore, conventional traffic monitoring systems face a significant challenge from an economical, monitoring range and time-response point of view because it requires to strategically install multiple sensors to cover monitoring area.

Distributed fiber-optic sensing (DFOS) based traffic monitoring systems utilize optical fiber cables, which are already laid along roads and deployed for communication purposes, can be integrated with the conventional traffic monitoring systems to improve the monitoring range and continuous data measurement *(3)-(5)*. DFOS systems measure acoustic vibrations, temperature, and strain, originating from the environment surrounding these fiber cables, by measuring change in Rayleigh backscattered signals in corresponding optical fiber cable sections to detect and localize vibrations *(6)-(9)*. This makes it possible to detect and localize traffic-induced vibrations along roads for traffic monitoring. The vehicle vibrations are analysed to estimate average traffic speeds and detect traffic events. Nonetheless, non-recurring events that lead to partial blocking of roads e.g. presence of obstructions on single lane of multi-lane highway, affect the traffic flow gradually in small measures. Therefore, DFOS systems that estimate average traffic speeds but do not consider vehicle lane-of-travel, face a challenge in detecting single-lane abnormalities.

However, DFOS systems have the capability to detect these uncertain events by monitoring vehicle behaviour along the monitoring sections. These non-recurring uncertain events, termed as single-lane abnormal events, cause an increase in spontaneous braking and avoidance driving maneuver. This leads to an increase in lane changes from lanes with obstructions and merging into unblocked lanes to avoid obstructions. Here, the vehicles desiring lane-change often induce large perturbations in traffic flow resulting in all-lane traffic congestion *(11)-(13)*. The time for congestion in such scenarios also depends upon the amount of traffic and thus, should be taken into consideration. Nevertheless, frequent lane changes, performed by multiple vehicles at the same section of the road, may indicate the occurrence of uncertain events. Hence, it is important to identify and locate these lane-change maneuvers as promptly as possible.

In this paper, we propose a technique to detect single-lane abnormal events by detecting these abnormal driving maneuvers with DFOS systems using a method to firstly track multiple vehicles on a road by adapting a particle tracking method to locate vehicle positions in the monitoring section; and secondly, to detect vehicle lane-change maneuver by monitoring changes in vehicle vibration characteristics for lane-change events by classifying lane-of-travel. We provide detailed explanations on the working of these methods in the following sections. Finally, we present the results for evaluation of the performance of the proposed methods for real traffic data collected along Shin-Tomei and Ken-O expressways in Japan.

**DFOS SYSTEMS FOR ABNORMAL EVENT DETECTION**

An illustration of a DFOS based traffic monitoring system consisting of an optical interrogator, connected to a fiber cable laid along a road section, that measures backscattered signals is shown in **Figure 1**. The upper part of **Figure 1** illustrates traffic congestion caused by an obstruction on both lanes of a two-lane road. The optical interrogator leverages the Rayleigh backscattering principle to sense and localize vehicle vibrations along the fiber cable. The lower part of **Figure 1** shows the corresponding vehicle vibrations on a time-distance plane, also known as a waterfall trace. The vehicles travelling on the lane closer to the fiber cable, termed as near-lane, produce higher intensities as compared to vehicles travelling on lane farther away from fiber cable *(14)*, termed as far-lane. The trajectories for vehicles moving in both directions (towards and away from optical interrogator) are shown in the waterfall trace. DFOS systems detect traffic anomalies by monitoring various traffic flow parameters such as speed, count, or vehicle density. These traffic parameters are calculated by analysing vehicle trajectory properties such as slope, count, thickness, or density. During congestion, vehicles slow down and the corresponding change in their vehicle trajectories, for e.g. change in slope (observed as vertical trajectories) indicate change in speed. Algorithms like TrafficNet denoise DFOS data by extracting useful vehicle trajectories and calculate an average traffic speed to detect traffic congestions *(15)*. Due to this, these solutions detect only large disruptions in traffic flow but cannot detect single-lane abnormal events because they estimate average traffic flow parameters without considering vehicle lane-of-travel. Thus, conventional DFOS systems face difficulties in quickly identifying abnormal events at the time of their occurrence, i.e. before a large disruption in traffic flow takes place.

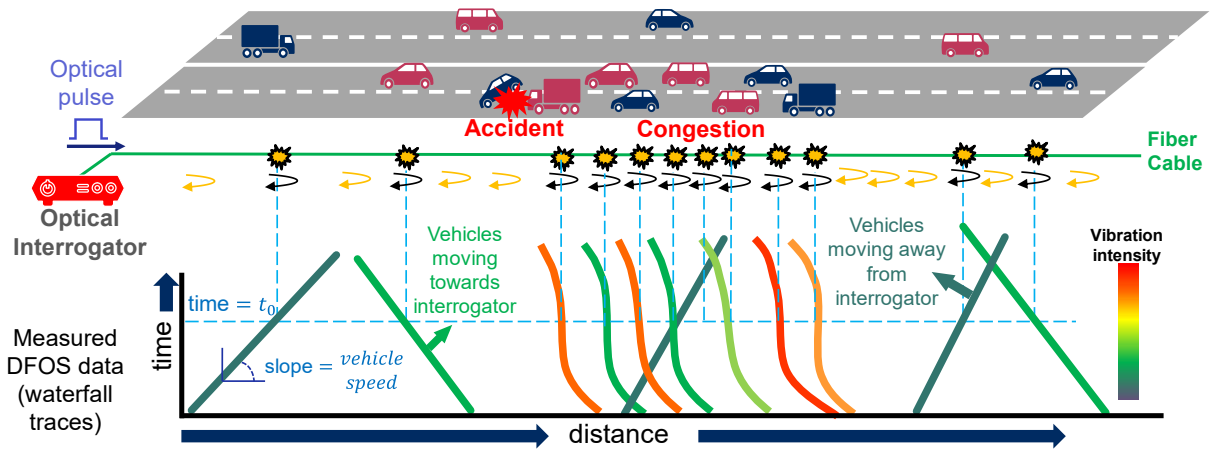

**Figure 1 Illustration of a typical distributed fiber-optic sensing-based traffic monitoring system**

**Related work**
Various traffic incident detection techniques utilize different types of sensors and incident detection algorithms *(16)(17)*. Among these, computer vision techniques to track individual vehicles travelling through intersections to detect accidents are increasingly being used *(18)(19)*. Here, traffic camera-based surveillance systems are strategically placed to confirm the existence of any incident and assess its severity. Nonetheless, the performance of these systems is affected by weather conditions, physical damage and the requirement of regular maintenance. These systems also face a challenge due to shorter monitoring range which makes it difficult to promptly detect uncertain events because of delay in propagation of congestion at the sensor location. DFOS systems can be integrated with these sensors to improve the sensing range and to reduce detection time. Field trials using DFOS systems utilizing deployed telecom fiber cables as sensing media with different network topologies have demonstrated usability in city traffic monitoring applications such as future road design, traffic signal planning, and for computing essential travel *(20)*. Furthermore, image segmentation-based AI models, like TrafficNet, have been trained and validated to extract vehicle trajectories, calculate average vehicle speeds, and monitor traffic flow in real noisy environment *(15)*. Nonetheless, DFOS solutions face difficulty in identifying abnormal events in low-density traffic flow i.e. a smaller number of vehicle trajectories to extract. Our motivation, in this work, is to propose an innovative solution to integrate the reliable conventional sensors, such as traffic cameras, with wide-area monitoring DFOS systems to promptly detect abnormal events before they cause significant traffic congestions.

*H. Prasad, Y. Yajima, D. Ikefuji, T. Suzuki, S. Tominaga, H. Sakurai, and M. Otani*

**EFFECT OF TRAFFIC FLOW ON ONSET OF TRAFFIC CONGESTION**

In real situations, traffic congestions occur not only because of lower road capacity but also due to abnormal vehicle driving behaviour such as changing lanes quickly or stopping suddenly in presence of abnormalities such as traffic accidents, vehicle breakdowns, or fallen debris in travel lanes *(21)*. Here, the amount of traffic flow plays a significant role to determine the severity of these abnormalities. When single-lane abnormalities occur in low-density traffic flow, no significant change due to spontaneous braking and abrupt lane-change maneuvers, occur in average traffic speed because of large headways. The headways decrease over time as vehicles begin changing lanes and merging into unblocked lanes more frequently. This frequent lane-changing eventually leads to traffic congestion in all lanes of the road. However, for a high-density traffic flow, time for onset of congestion is comparatively shorter because vehicle headways at the occurrence of abnormalities are smaller and thus, result in congestion more quickly. We simulate a single-lane abnormal scenario to understand and validate this effect of amount of traffic flow on onset of traffic congestion. The motivation for this simulation is to understand traffic flow conditions for which prompt incident detection may be difficult. We implement a cellular automata (CA) model based stochastic Nishinari-Fukui-Schadschneider (S-NFS) model *(22)* to determine the effect of traffic flow on traffic congestion due to spontaneous-braking and lane-changing in presence of obstructions on a single-lane.

**TABLE 1 A list of parameters for single-lane abnormal event simulation**

| Parameter | Value Choice |
|---|---|
|  |  |
| **Highway parameter** |  |
| Number of lanes | 2 |
| Length of highway | 5km |
| Maximum speed | 100km/h |
| Obstruction location | Slow lane |
| Obstruction distance | 4 to 4.1km (100m) |
| Speed in obstruction location | 0km/h |
|  |  |
| **CA model parameter** |  |
| Random braking probability | 0.3 |
| Lane change probability | 0.05 |
| Slow-to-start probability | 0.1 |
| Anticipation probability | 0.98 |
| Pass lane probability | 0.6 |
| Speed limit | 80km/h |
| Traffic in-flow | 3 to 35 vehicles/min |

**Simulation condition**

CA models can decompose a complex phenomenon into a finite number of elementary processes using simple approaches to model and simulate dynamic systems. Due to this, CA models are used to model vehicular traffic flow, collective behaviour, and self-organization to simulate congestions on freeways. In this paper, we adopt a CA-based S-NFS model (with model parameters explained in **Table 1**) to simulate traffic congestion scenarios for different traffic flows. The simulation conditions are mentioned in **Table 1** and explained in **Figure 2**. The mentioned CA model parameters are chosen to simulate real highway traffic flow. Here, the obstruction section simulates vehicles changing lanes to avoid obstructions and the average traffic speeds for every 30secs in every 100m of the highway sections for the simulated traffic flow are

*H. Prasad, Y. Yajima, D. Ikefuji, T. Suzuki, S. Tominaga, H. Sakurai, and M. Otani*

estimated. The estimated time series average speed data, shown in **Figure 2** (left), is then used to determine the time for onset of traffic congestion, where the start of traffic congestion is defined as an instance when the estimated average traffic speed is continuously lower than 40km/h for more than 1min and for highway section longer than 1km. This definition of traffic congestions is commonly used for road management in Japan. To evaluate the effect of traffic flow on onset for traffic congestion in presence of single-lane abnormalities, we estimate the time taken for the onset of congestion from the time of occurrence of event on single-lane of the simulated two-lane road section, explained in **Figure 2** (left). We consider different average traffic flows on the highway by changing the traffic in-flow parameter accordingly. Here, the traffic in-flow is randomized between two consecutive traffic flow values for a more realistic simulation.

**Simulation results [traffic flow v/s time for onset of congestion]**
The simulation result in **Figure 2** (right) shows a trend that represents time for congestion in a traffic scenario for different traffic flow. We also show different traffic flows for two expressways, Ken-O [~05:00AM], Ken-O [~05:00PM] and Shin-Tomei [~12:30PM] (also considered for evaluation in this paper) with average traffic flows 5-6, 34-35 and 14-15 vehicles/min respectively. Thus, traffic flow depends not only on the type of road but also time of the day. Here, the estimated time for onset of congestion for Ken-O and Shin-Tomei expressways may vary for different expressways and time of the day. We observe that the time taken for a congestion to occur for the simulated scenarios is longer for low-density traffic flow and shorter for high-density traffic flow, for example, around 9 mins for 3-4 vehicles/min and less than 1 min for 23-24 vehicles/min and beyond. Here, we also observe that the time for onset of congestion decreases gradually with increase in traffic flow. A sudden fall in time for onset of congestion can be observed around 21-22 vehicles/min. This suggests that beyond this traffic flow boundary, abnormal events can be detected immediately after their occurrence using conventional DFOS based traffic flow monitoring techniques that employ average traffic speed estimation. The results of this simulation show that for traffic scenarios with low-density traffic flow, i.e. below the traffic flow boundary, there is a need for methods that detect the occurrence of abnormalities faster than the onset of traffic congestions.

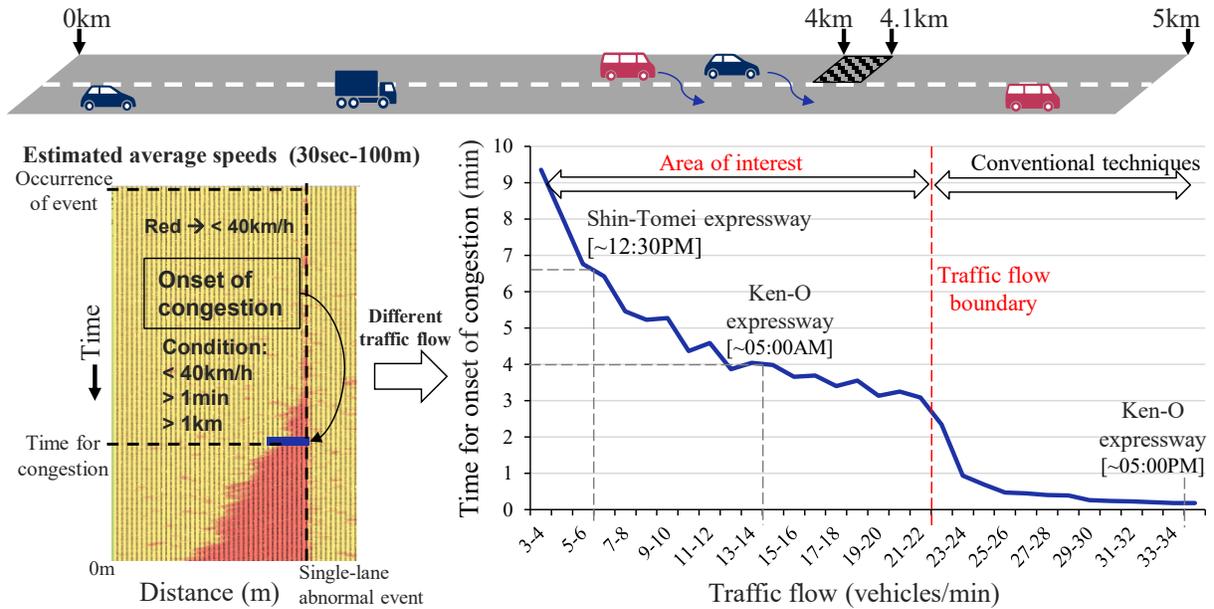

**Figure 2 Simulation results to evaluate effect of traffic flow on onset of traffic congestion for a two-lane road section with single-lane abnormal event.**

H. Prasad, Y. Yajima, D. Ikefuji, T. Suzuki, S. Tominaga, H. Sakurai, and M. Otani

**PROPOSED METHOD FOR SINGLE-LANE ABNORMAL EVENT DETECTION**

In this paper, we propose methods to identify and locate occurrence of single lane abnormalities on a two-lane highway in low-density traffic flow. To do so, we estimate individual vehicle positions and then detect abnormal vehicle maneuvers. First, we propose a method to track a vehicle along a road section by localizing vehicle positions along their respective path in DFOS data. Next, we propose a method to identify lane-change maneuver of the tracked vehicles along a road section by monitoring variations in vehicle vibration intensities and frequency along the estimated vehicle paths. These two methods are performed iteratively in a monitoring section. Furthermore, the output of lane-change detection can be statistically analysed to determine the frequency and propagation of lane-change events. In this paper we focus on evaluating the performances of individual vehicle tracking and lane-change event detection.

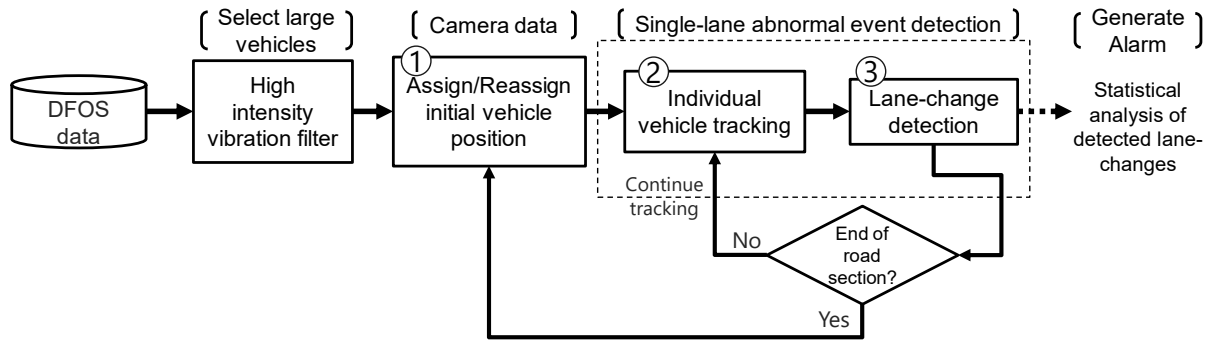

Figure 3 Overall block diagram of proposed single-lane abnormal event detection method.

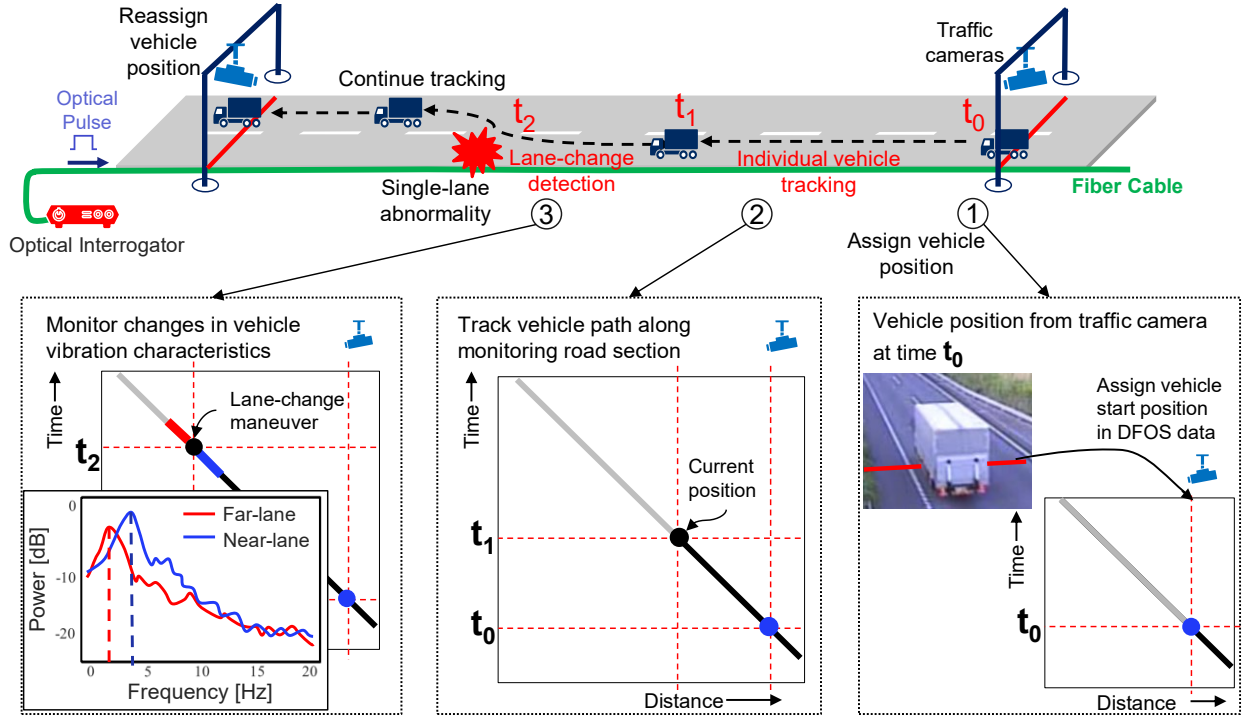

Figure 4 Schematic diagram of single-lane abnormal event detection with a vehicle travelling in the monitoring section (between camera positions) that changes lanes to avoid a single-lane obstacle.

The overall block diagram of our proposed methods for tracking individual vehicles and monitoring vehicle behaviour to detect single-lane abnormalities in low-density traffic flow is explained in **Figure 3**. For a more robust and reliable performance, vehicles producing high vibration intensities such as trucks were considered in this work. This is because in DFOS data, monitoring sections may produce low signal-to-noise ratio (SNR) due to the structural noise and inconsistent fiber layout conditions. These low SNR sections lead to large discontinuities in vehicle trajectories and induce noise in measured data. Therefore, the data is first pre-processed with a suitable intensity-filter to supress low-intensity vehicle trajectories such as cars. Next, the vehicle is located at the camera position and vehicle seed-point is assigned to respective vehicle trajectory in DFOS data. After this, we iteratively perform individual vehicle tracking and lane-change detection in the monitoring section. The detected lane-change maneuvers can then be analysed by statistical approaches to determine presence of any abnormal events. For the vehicles that reach the end of the monitoring sections, the seed-point is reassigned to the vehicle trajectory for monitoring in the next section. The overall concept of the proposed method is explained in **Figure 4**. Here, the target vehicle enters the monitoring section near the start-camera position at time $t_0$ and travels on the near-lane. The vehicle seed-point is assigned to the respective DFOS vehicle trajectory in waterfall trace for same time and location, as illustrated in step 1 (right). The vehicle path is then traced iteratively, and the current vehicle position is continuously updated at time $t_1$ as illustrated in step 2 (center). The vehicle avoids a single-lane abnormal event and changes lanes from near to far-lane of the highway at time $t_2$. The change in vehicle characteristics on near and far-lane are illustrated in step 3 (left). A shift in spectral centroid of vehicle vibration characteristics due to change in lane-of-travel can be observed as vehicle vibration intensity decrease in far-lane. When the vehicle reaches the end of the monitoring section, the vehicle path is confirmed, and a new seed-point is assigned for the vehicle trajectory for the next monitoring section.

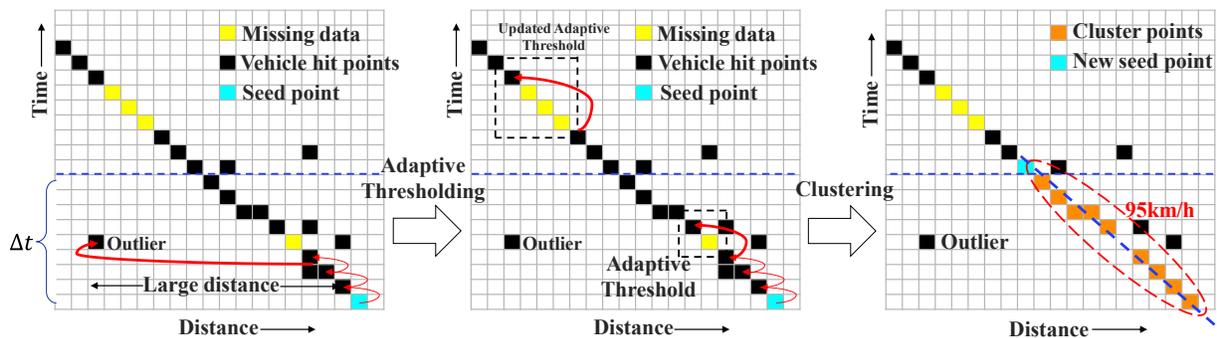

**Figure 5 An illustration of vehicle hit point de-noising technique to estimate vehicle path in DFOS.**

**Individual vehicle tracking**
In our proposed method, we locate the positions of vehicles travelling along the monitoring section, i.e., between two traffic camera positions. Here, we adapt a particle tracking technique, used for tracking charged particles in accelerators, to identify and trace vehicle path in a road section *(23)*. In a particle accelerator, the positions of moving particles, originating from a seed point, are detected along many detector layers for different time instances and a track is constructed using this collection of particle positions. Similarly, in DFOS data, the vehicle positions (with seed point at a camera location) can be located at different time instances along the fiber and then, a vehicle path can be constructed using these vehicle positions. First, we localize the vehicle start positions observed in traffic camera as the seed-point for corresponding vehicle path in DFOS data. We then detect the peak vehicle vibrations intensities at different time instances. Here, the peak vibrations represent vehicle positions along the given road section

H. Prasad, Y. Yajima, D. Ikefuji, T. Suzuki, S. Tominaga, H. Sakurai, and M. Otani*(24)* and are termed as vehicle-hit points. The propagation of these vehicle-hit points is observed as the vehicle trajectories or path of the vehicle along the fiber cable in DFOS data. We then construct a path of travel using these vehicle-hit points for a short duration of time as explained in **Figure 5**. We first estimate all vehicle hit points for a short interval ($\Delta t$) as illustrated in **Figure 5** (right). We then find the closest hit points at all time instances for $\Delta t$ from the seed-point as explained in **Figure 5** (center). Noisy vibrations in DFOS data, due to structural noise, fiber layout, or ambient background noise, may lead to estimation of noisy peak vibrations as vehicle-hit points. We eliminate any noisy hit points by performing an adaptive thresholding, based on estimated vehicle speed and physical limitations, on the distance and time of travel, for the moving vehicles as shown in **Figure 5** (center). This means that vehicle positions outside this threshold are not achievable for this vehicle. The adaptive threshold also checks for missing data points and updates the threshold range to locate the next hit point as explained in **Figure 5** (center). Next, we perform an unsupervised clustering on these denoised datapoints using K-Means clustering, with time and position as parameters, to fit a vehicle path among as illustrated in **Figure 5** (right) and explained in **Equation 1**.

$$J = \sum_{i=1}^{p} \sum_{j=1}^{k} w_{ij} \left\| h_j^{(i)} - c_i \right\|^2, \qquad (1)$$

where $J$ is the objective function, $k$ is total number of vehicle-hit points, $p$ is total number of vehicles, $i$ is cluster number, $j$ is number of vehicle-hit points, $h$ denotes vehicle-hit point, $c_i$ is centroid of cluster $i$, $\|h_j^{(i)} - c_i\|^2$ is Euclidean distance function and $w_{ij}$ is '1' when vehicle hit point $x^i$ belongs to cluster $k$; and '0' otherwise. The goal is to minimize the objective function w.r.t $c_i$ and $w_{ij}$. Here, the datapoints are assigned to initial centroids using Euclidean distance and new cluster centroids are computed for minimised cluster variance. We chose K-Means clustering for a faster and simpler approach for constructing the vehicle-hit points. It is possible to adopt more complex approaches with additional parameters such as vibration intensity, trajectory thickness or vibration frequency to construct the vehicle path. The hit points, estimated as part of vehicle trajectories, are then used to estimate vehicle parameter such as vehicle speed Vehicle speed is then used to estimate the new speed point for next set of data and hence, vehicle tracking is performed iteratively. Here, a least squares method, as explained in **Equation 2**, is used to determine slope of a line of best fit for the vehicle path for clustered datapoints. We then estimate the vehicle speed using spatial and temporal resolutions of measured DFOS data.

$$m = \frac{\sum_{u=1}^{n}(x_u - \bar{x})(y_u - \bar{y})}{\sum_{u=1}^{n}(x_u - \bar{x})^2}, \qquad (2)$$

where m is slope of best fit line for cluster datapoints, $n$ is total number of cluster points, $u$ is cluster point number, $\bar{x}$ is the mean of all x-coordinates and $\bar{y}$ is the mean of all y-coordinates.

**Lane-change detection**
We explain a method for detecting lane-change maneuver of the tracked vehicles in this section. The measured vibration intensities and frequency characteristics of a vehicle change with change in distance of the source vibration from the fiber cable due to distance attenuation *(14)*. Moreover, the effect on higher frequency components due to distance attenuation is larger. In the other words, vehicle spectral centroid, which indicates the center of mass of vehicle vibration, depends on the lane-of-travel. **Figure 6** demonstrates a comparison between corresponding amplitude, spectrogram, and spectral centroid of vehicle vibrations on near and far-lane for a vehicle performing lane-change maneuver on an expressway. Here, spectral centroid at each position $SC(l)$ is derived from **Equation 3**.

H. Prasad, Y. Yajima, D. Ikefuji, T. Suzuki, S. Tominaga, H. Sakurai, and M. Otani

$$SC(l) = \sum_{l-r}^{l+r} \frac{\sum_{f=0}^{F-1} f|S(l,f)|}{\sum_{f=0}^{F-1} |S(l,f)|}, \qquad (3)$$

where, $S(l,f)$ indicates a spectrum at position $l$ with frequency $f$, $F$ is the maximum frequency for deriving a spectral centroid and $r$ indicates the range of position for calculating an average spectral centroid. An averaging along the distance is performed to reduce fluctuations in calculated spectral centroid. Here, $r$ is defined to average over a range of 40m. From **Figure 6**, we observe that the spectral centroid is stable in both lanes as compared with the amplitude of vehicle vibration. Therefore, the spectral centroid is employed for lane-change detection in this paper. Here, the spectral centroid depends on the type of vehicle type and road structure. To reduce the influence of these dependencies, the proposed method measures statistical spectral centroid of vehicle vibrations for each lane of the monitoring expressway beforehand. The statistical spectral centroid is utilized for defining the threshold frequency to classify lane-of-travel.

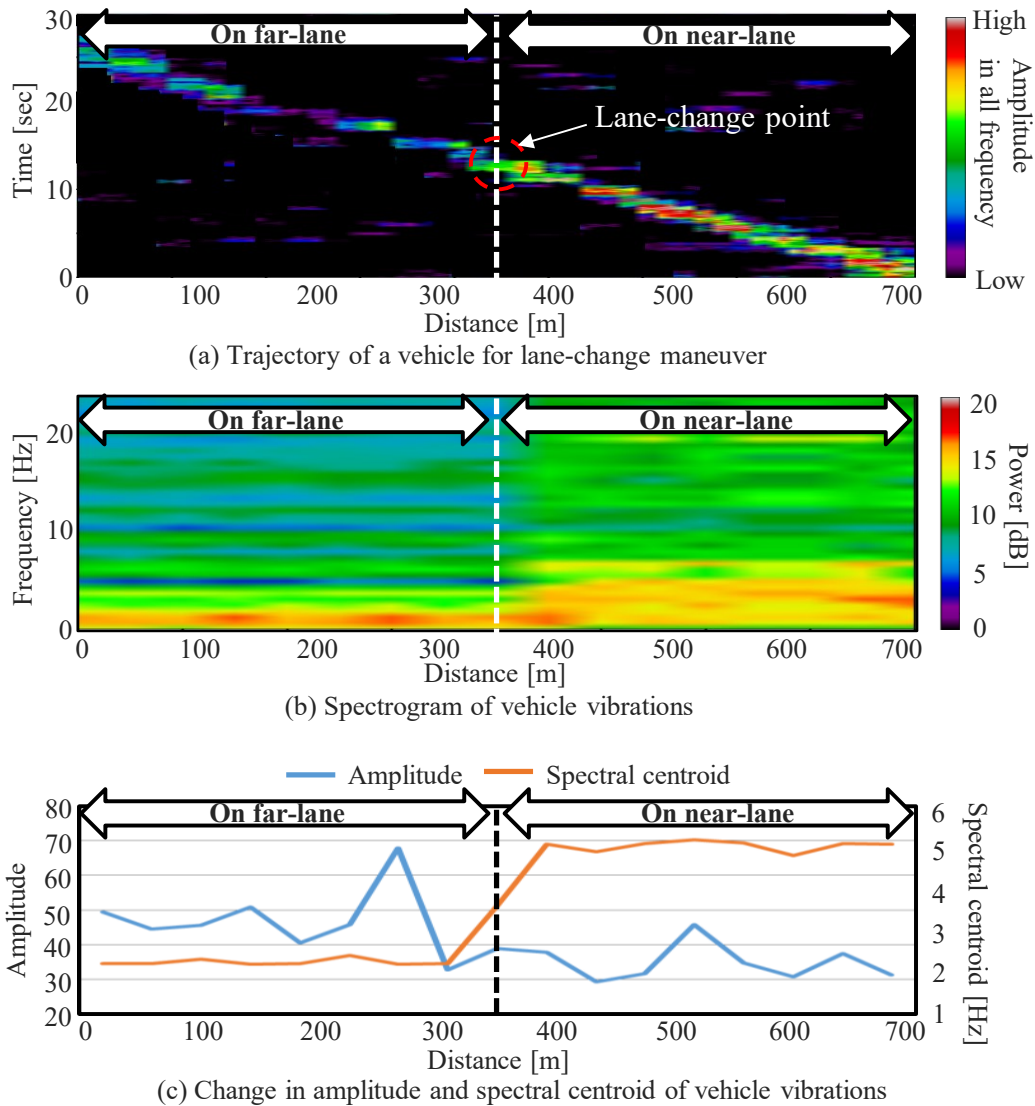

(a) Trajectory of a vehicle for lane-change maneuver

(b) Spectrogram of vehicle vibrations

(c) Change in amplitude and spectral centroid of vehicle vibrations

**Figure 6 Comparison of (a)vehicle trajectory, (b)spectrogram, and (c)change in amplitude and spectral centroid of vehicle vibrations on near and far-lane for lane-change maneuver.**

*H. Prasad, Y. Yajima, D. Ikefuji, T. Suzuki, S. Tominaga, H. Sakurai, and M. Otani*

**EXPERIMENTS**

We conducted field trials at two expressways in the Kanagawa prefecture of Japan to evaluate the performance of our proposed methods for single-lane abnormal event detection. An existing fiber-optic cable infrastructure alongside these expressways, provided by Central Nippon Expressway Company (C-NEXCO), was used for distributed sensing using an optical interrogator. The fiber cables for both these expressways are laid along the shoulder of the expressways in one direction. We measured DFOS data in road sections (between two traffic camera positions) for Ken-O [05:00:00 - 05:20:00] and Shin-Tomei [12:20:00 - 13:15:00] expressways with temporal resolutions of 1msec and spatial resolutions of 3.2m and 8.16m, respectively. We down sample the data to 0.2sec temporal resolution for vehicle tracking to reduce computation and effect of noise. We consider two different expressways to validate the robustness and scalability of these methods for different traffic flow and fiber layout conditions. The traffic flow for Ken-O and Shin-Tomei expressways were about 14-15 and 5-6 vehicles/min respectively for the measured data. For Ken-O expressway, we use a road section of length 1.2km between 20.3KP (start-camera) and 19.1KP (end-camera). Similarly, for Shin-Tomei expressway, we use a road section of length approximately 630m between camera positions 20.2KP (start-camera) and 19.57KP (end-camera). Here, kilo-posts or KP is an indicator used by Japanese road operators to indicate distance (in unit km) from the starting point of a highway. We use the traffic cameras, along these expressways, to identify the seed-point and to validate the performance of individual vehicle tracking by matching the end-position of the vehicle paths estimated in DFOS and traffic cameras. We also use camera data to locate lane-changing vehicles, either from far to near lane or vice versa, in the camera field-of-view as reference events for our evaluation.

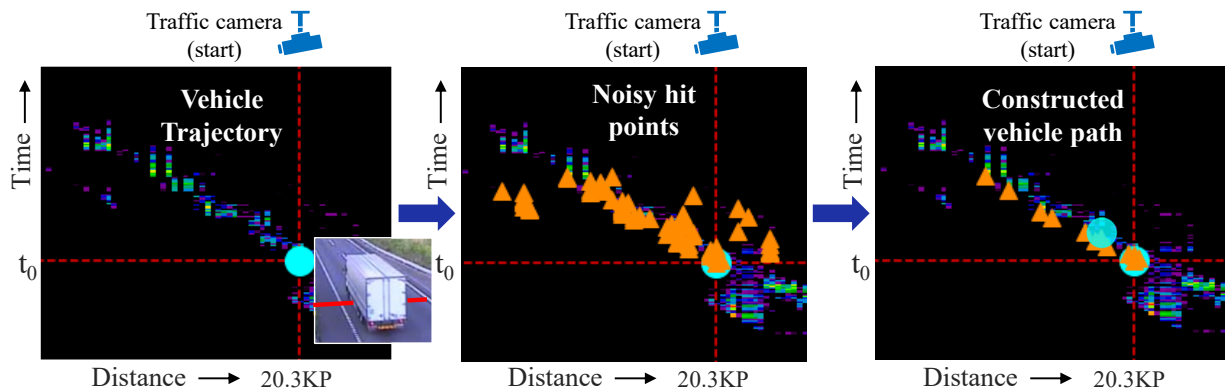

**Figure 7 Vehicle path tracing using hit point estimation (left), dynamic thresholding (center) and vehicle path clustering (right). (white colour)**

**Individual vehicle tracking evaluation**

The start-time of a vehicle ($t_0$) (from start-camera) and camera position are used to assign vehicle seed-point for each vehicle path in DFOS data. All peak vehicle intensities as vehicle-hit points, in a 10sec time window for every time instance (0.2sec interval), are estimated to fit a vehicle path and to calculate the vehicle speed. Then, we calculate the new position of the vehicle after 1sec, the next seed-point, calculated for vehicle path construction in next 10sec window. **Figure 7** shows the steps for path construction a vehicle on Ken-O expressway for 10sec time window. Here, the seed-point observed from traffic camera (left), all vehicle-hit points with noisy hit points (center) and constructed vehicle path with updated seed point (right) are shown. Here, erroneous hit points are estimated due to structural noise (bridge structure, fiber layout) or low SNR. Nevertheless, a vehicle path for the respective vehicle trajectory in DFOS can be constructed and seed-point for next 10sec window is calculated. We iteratively construct vehicle paths at all time instances for the vehicle travelling inside the monitoring section (between two cameras) to track this vehicle and monitor vehicle behaviour. **Figure 8** shows the constructed vehicle paths of three vehicles in DFOS waterfall trace with their respective camera snapshots at start and end-camera positions. The traced vehicle paths follow their respective trajectories even when there are missing data sections due to low SNR.

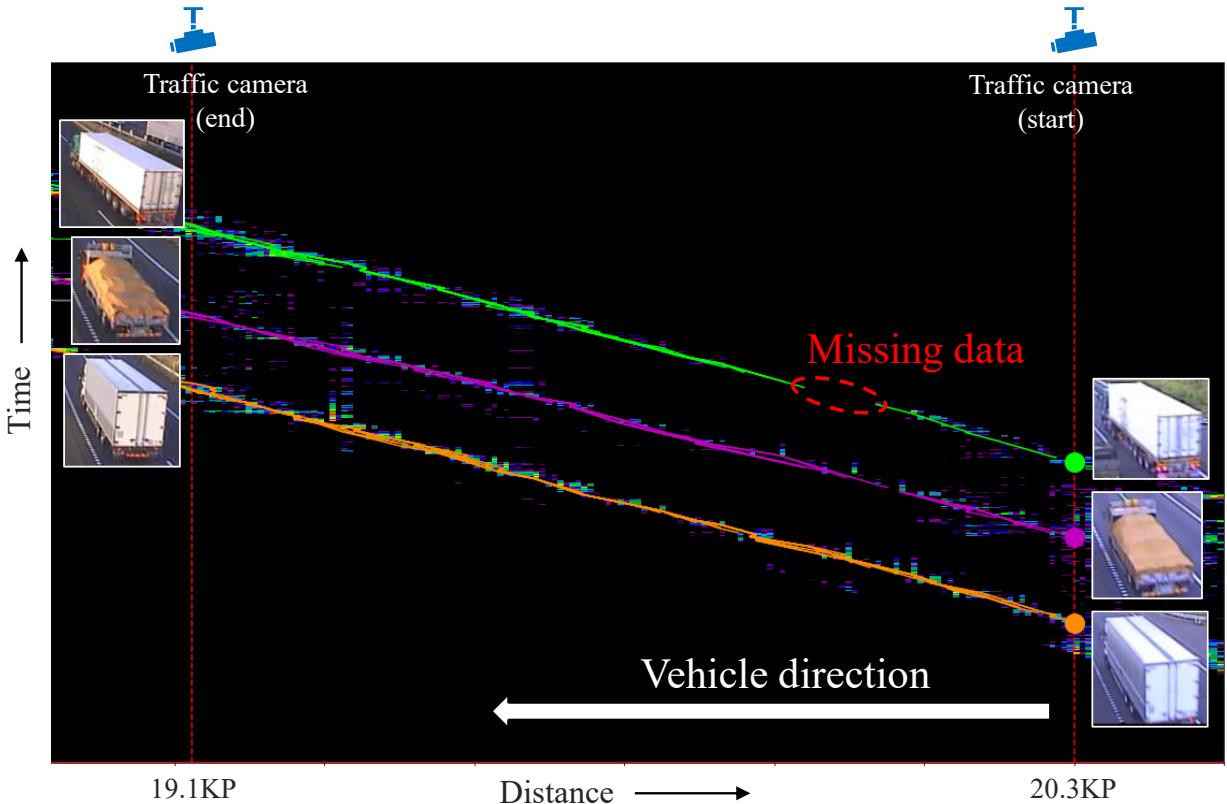

**Figure 8 Estimated vehicle trajectories for different vehicles travelling on Ken-O expressway between 20.3 KP and 19.1 KP**

*Individual vehicle tracking performance*
We validate the scalability and robustness of our proposed vehicle tracking method for two different expressway sections with different traffic flows. In DFOS, sensitivity decreases as the distance from the start of cable increases due to fiber insertion loss and weak backscattering signals. Also, SNR for each highway section may change depending on the highway structures and fiber layout. Therefore, we consider vehicles producing large vibrations, such as trucks and tankers, for robustness. The tracking accuracies for Ken-O and Shin-Tomei expressway are shown in **Table 2**. We calculate the accuracy of individual vehicle tracking using the following expression:

$$Individual\ vehicle\ tracking\ accuracy\ (\%) = \frac{number\ of\ correct\ vehicle\ paths}{total\ number\ of\ vehicles} \times 100$$

where, $number\ of\ correct\ vehicle\ paths$ denote vehicles whose path were estimated correctly along their respective vehicle trajectory and confirmed at end-camera position and $total\ number\ of\ vehicles$ is the number of truck paths for tracking. The validation results show an average accuracy of about 81.5% for considerable number of vehicles travelling on different highway sections.

**TABLE 2 Individual vehicle tracking performance for large vehicles.**

|  | Number of correct vehicle paths | Total number of vehicles | Tracking accuracy |
|---|---|---|---|
| **Ken-O Expressway** | 75 | 94 | 80% |
| **Shin-Tomei Expressway** | 22 | 25 | 88% |
| **All vehicles** | 97 | 119 | 81.5% |

**Lane-change detection analysis**

In our evaluation, we use lane-change maneuvers, performed by different vehicles in the field-of-view of traffic cameras, installed on the Ken-O expressway, as reference events. Due to the restrictions in collecting data for a real single-lane abnormal scenario in a safe and economical manner, we consider the multiple vehicles performing lane-changes at the camera positions to be same as multiple vehicles performing lane-change maneuver during single-lane abnormal events. In our evaluation, we collected vehicle vibrations for lane-change maneuver using data from five camera locations for 1 hour as reference. For Ken-O expressway, we observed a total of 40 lane-change events for vehicles changing lanes from near to far lane and 37 events from far to near lane. Also, to evaluate the scalability of this method on different road sections, we performed and measured lane-change events by driving a dash-camera mounted heavy vehicle on Shin-Tomei expressway. Next, we identified these lane-change points for respective time and location in DFOS data. For Shin-Tomei expressway, the reference lane-change events, observed from dash-camera data, were 30 lane-change events for both lane-change scenarios (near to far lane and vice versa).

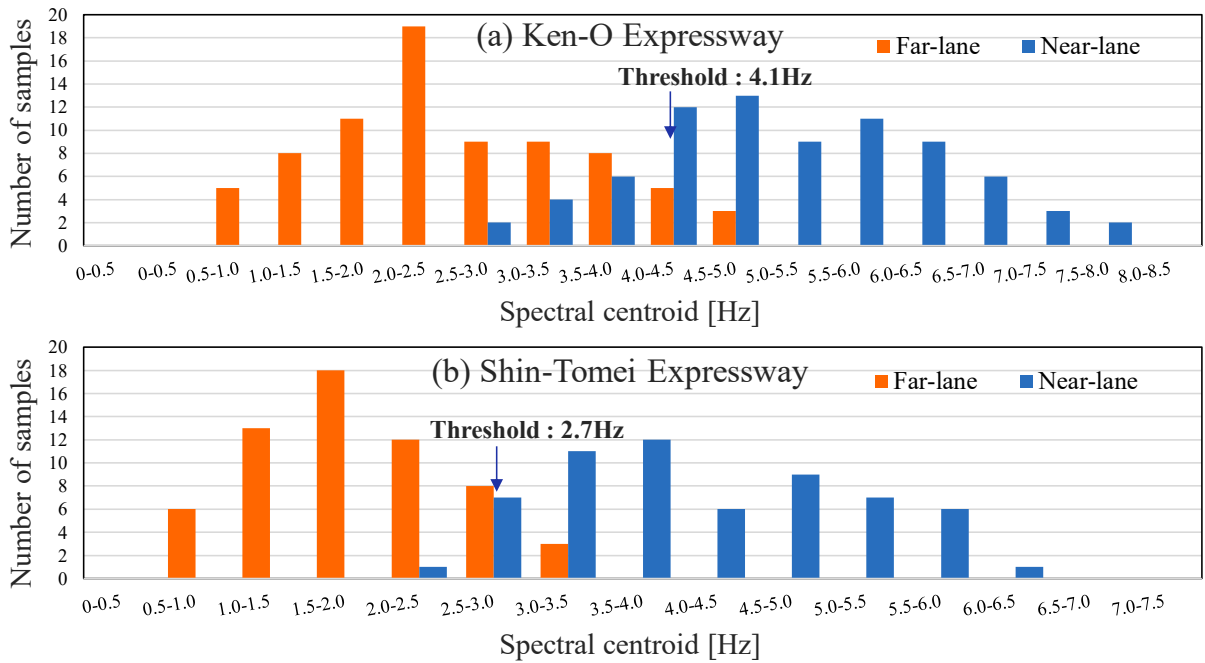

**Figure 9 Distribution of spectral centroids of vehicle vibrations on both lanes (before and after lane-change point) for Ken-O expressway (top) with threshold frequency at 3.3Hz and Shin-Tomei expressway (bottom) at 2.7Hz.**

*Lane-change detection performance*
To confirm the performance of the lane-change detection method for Ken-O and Shin-Tomei expressways, we identified lane-change points on DFOS data for all observed events and performed lane-change detection. The vehicles change lanes for two scenarios, either from near to far-lane and vice versa. We calculate the spectral centroids for all vehicle vibrations before and after lane-change points for both lane-change scenarios. The histogram of estimated spectral centroids for vehicles on near and far-lane is shown in **Figure 9**. Here, the bi-modal distribution can be threshold at 4.1Hz frequency, for Ken-O expressway, to classify the vehicle lane-of-travel. The vehicle vibrations with spectral centroid below 4.1Hz are travelling on the far-lane and above 4.1Hz are travelling on the near-lane. Using this threshold frequency, we detected lane-change events by classifying lane-of-travel for vehicle vibrations before and after lane-change point. Similarly, for Shin-Tomei expressway, the threshold frequency was estimated at 2.7Hz and lane-change detection was performed. **Table 3** shows the performance of lane-change detection for vehicles by classifying lane-of-travel before and after lane-change point. The average accuracy of lane-change detection for Ken-O and Shin-Tomei for both lane-change scenarios are 85% and 82% respectively where lane-of-travel before and after lane-change point were correctly classified. We calculate the accuracy of individual vehicle tracking using the expression below:

$$Lane\ change\ detection\ rate\ (\%) = \frac{number\ of\ detected\ lane\ change\ events}{total\ number\ of\ lane\ change\ events} \times 100$$

**TABLE 3 Lane-change detection performance considering initial lane of travel.**

|  | Lane-change scenario | Number of detected lane-change events | Total number of lane-change events | Detection rate |
|---|---|---|---|---|
| **Ken-O Expressway** | Near to far-lane | 34 | 40 | 85% |
|  | Far to near-lane | 32 | 37 | 84% |
| **Shin-Tomei Expressway** | Near to far-lane | 26 | 30 | 87% |
|  | Far to near-lane | 23 | 30 | 77% |
| **All vehicles** | Near to far-lane | 60 | 70 | 85% |
|  | Far to near-lane | 55 | 67 | 82% |

**DISCUSSIONS**

**Abnormal event detection for multi-lane highways**
In this work, we demonstrated and evaluated a single-lane abnormal event detection method for two-lane roads. Nevertheless, this idea can further be implemented to detect abnormal events on highways with two or more lanes with blocked lanes. This is because the tendency of vehicles to perform lane-change maneuver does not depend upon the number of blocked lanes. The proposed method can then be used to detect vehicle changing lanes due to obstacles partially blocking the highways, i.e. blocking traffic flow on 1 or more lanes of multi-lane highways.

**On-ramp merging support**
The proposed method can be implemented for automated on-ramp merging support in autonomous vehicle driving. The position and lane-of-travel of non-connected vehicles on highway sections (main road or on-ramp) can be shared with connected vehicles for an efficient automated and cooperative on-ramp merging control. This can ensure safety of vehicle driving and reduce the risk of collisions at the merging zones.

**Estimation of vehicle path in noisy sections**
The error in vehicle tracking arises due to incorrect vehicle path estimation, where the constructed path does not belong to the respective vehicle trajectory. This is caused by noise in low SNR sections when multiple vehicles travel closely together (small headway) in crowded scenarios such as platooning or overtaking. This leads to estimation of vehicle paths repeated over the same vehicle trajectory and thus, leading to no vehicle path estimation for the missed vehicle trajectories as explained in **Figure 10**. To improve the vehicle tracking performance, road sections with high SNR with clear vehicle trajectories even in crowded scenario in corresponding DFOS data should be considered.

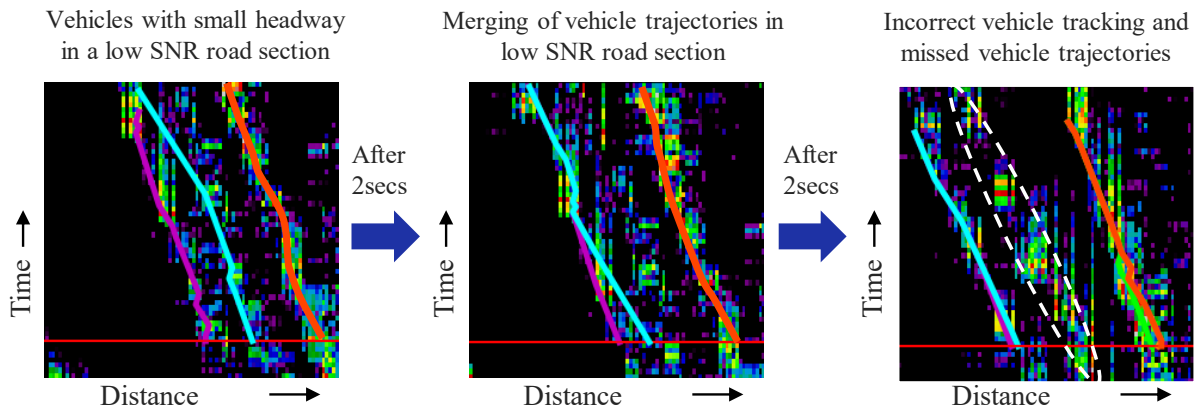

**Figure 10 Issues with individual vehicle tracking for crowded travelling scenario.**

**Estimation of different vehicle characteristics**
Various vehicle characteristics such as thickness or amplitude along constructed vehicle paths can be estimated to provide additional information related to weight of passing vehicles. The vehicles identified using camera systems cannot provide the vehicle weight information. Vehicle vibrations also depend on the weight of vehicles. DFOS can be used to identify overweight vehicles at respective camera positions by comparing vehicle size and vibration characteristics.

**CONCLUSIONS**
In this work, we presented and validated methods using DFOS systems for identifying the presence of single-lane abnormalities on highway sections. Our motivation was to detect non-recurring traffic congestion that is often uncertain and relatively difficult to locate and identify. We first proposed a method to identify vehicle positions in DFOS data using camera data, to locate vehicle seed-point, and construct vehicle paths in the monitoring sections. We also proposed a method to detect and locate lane-change maneuvers performed to avoid single-lane obstructions in abnormal events. We illustrated the proposed methods' capabilities to overcome background noise and data gaps in DFOS data for vehicle path tracking and lane-change detection. Finally, we evaluated the performance of our proposed methods by conducting a field trial along road sections on Ken-O and Shin-Tomei expressways in Kanagawa, Japan. The average error in vehicle path tracking of 119 trucks was around 18.5% and for vehicle lane-change detection was around 16.5%. These results suggest that the proposed method has potential for detecting occurrence of single-lane abnormal events in low-density traffic flow so that necessary mitigation measures can be implemented before onset of traffic congestions. Our future work includes combining the two proposed methods to automatically detect single-lane abnormalities on all highway sections and validating the performance with real abnormal event data. We will also focus on integrating an AI-based camera system to automatically classify vehicles to assign seed-points in DFOS data for continuous monitoring.

**AUTHOR CONTRIBUTIONS**

The authors confirm contribution to the paper as follows: study conception and design: H. Prasad, Y. Yajima, D. Ikefuji, S. Tominaga, H. Sakurai; data collection: D. Ikefuji, H. Sakurai, S. Tominaga, T. Suzuki, M. Otani; analysis and interpretation of results: H. Prasad, Y. Yajima, D. Ikefuji, H. Sakurai; draft manuscript preparation: H. Prasad, Y. Yajima, D. Ikefuji, H. Sakurai, M. Otani. All authors reviewed the results and approved the final version of the manuscript.